# Electro-optic and dielectric properties of Hafnium-doped congruent lithium–niobate crystals

Mustapha Abarkan*, Michel Aillerie ✉, Jean Paul Salvestrini, Marc D. Fontana

*Laboratoire Matériaux Optiques, Photonique et Systèmes, Université Paul Verlaine - Metz and Supélec, UMR CNRS 7132, 2 rue E. Belin, 57070 Metz, France*

Edvard P. Kokanyan

*Institute for Physical Research, National Academy of Sciences of Armenia, 0203, Ashtarak-2, Armenia*

* *permanent address: Faculté polydisciplinaire de Taza, Université Sidi Mohammed Ben Abdallah de Fès, BP 1223 Taza-Gare, Morocco*

**PACS numbers:** 78.20.J  77.22  77.84

## ABSTRACT

We report the optical and dielectric properties in hafnium (Hf)-doped lithium niobate (LN) crystals. We investigated samples of congruent composition with various doping concentration varying from 0 to 8 mol%. The clamped and unclamped values of the electro-optic coefficient $r_{22}$ of Hf-doped LN and the corresponding dielectric permittivity as well, have been experimentally determined and compared with the results obtained in undoped congruent LN crystals. We show that the electro-optic and dielectric properties are only slightly affected by the introduction of hafnium ions and therefore Hf-doped LN has the advantage of low photorefractive damage if compared with undoped congruent LN.

✉ Fax: +33-387-378559, E-mail: aillerie@metz.supelec.fr

## INTRODUCTION

Thanks to its excellent piezo-electric, electro-optical, and nonlinear optical properties, lithium niobate, $LiNbO_3$ (LN) has a wide range of applications in electro-optic (EO) modulation and laser Q-switching [1-4]. LN is not hygroscopic and offers good transmission and high extinction ratio with a modest driving voltage in the transverse configuration, i.e. when an electric field is applied perpendicular to the direction of the optical beam. Lithium niobate crystals exist in a wide range of composition from congruent (Xc=[Li]/[Li]+[Nb]= 48,6mol %) to stoichiometric ones (Xc=50 mol %). Congruent crystals present a large amount of non-stoichiometric defects and, due to its complex defect structure, it can accept a wide variety of dopants with various concentrations. Usually grown in congruent composition, undoped LN crystal suffers from a relatively low optical damage threshold, which constitutes the major drawback for LN based-optoelectronic devices, even for low power laser-beam applications [5]. It was proved that the optical damage threshold is dependent on the amount of intrinsic defects, and is considerably increased in stoichiometric LN and congruent LN doped with specific metal ions, such as Mg, Zn and In [6-7]. Recently, hafnium (Hf) was found to be a new optical damage-resistant element leading to a significant increase of the photorefractive resistance at doping threshold concentration between 2 and 3mol% [8-12].

It is therefore of interest for EO modulation or Q-switching applications, to know the EO coefficients values of hafnium- doped lithium niobate crystals as function of concentration over a wide frequency range from DC up to high frequencies. This knowledge is so as important since, in addition to photoconductivity, photorefractive behaviour is linked to the EO properties.

It is reminded that the high-frequency or constant-strain (clamped) coefficient $r^S$ is the true EO coefficient due to the direct modulation of the indices by the applied electric field, in contrast to the low-frequency or constant-stress (unclamped) coefficient $r^T$, which in addition includes the contribution of the lattice deformation by the electric field. As a consequence we have carried out EO measurements as function of frequency over a wide range and we report for the first time both clamped and unclamped EO coefficients values versus Hf concentration.

Here we investigated the EO coefficients $r_{22}^T$ and $r_{22}^S$ mainly involved in Q-switch laser applications. The coefficient $r_{22}$ is obtained when a light-beam is propagating along the optical axis (c-axis) of the crystal and it is therefore unaffected by the temperature dependence of the birefringence within this configuration. To complete the study, we have measured the frequency dependence of the corresponding dielectric permittivity $\varepsilon_{22}$ .

It should be mentioned that the EO coefficients $r_{13}^T$, $r_{33}^T$ and $r_c^T$ versus Hf concentration have been reported earlier by Rossella et al [13]. These coefficients were also determined within our study and the measured values were compared with those of ref [13]. The high-frequency coefficient $r_c^S$ and the corresponding acoustic contribution to the true EO coefficient $r_c$ were additionally given.

## EXPERIMENTAL

### Sample preparation

A set of Hf-doped lithium niobate crystals was grown by the Czochralski technique utilizing a growth set-up with rf heating element in air atmosphere using a single platinum crucible. The starting materials used for sintering the lithium niobate charges of congruent composition were high purity $Nb_2O_5$ and $LiCO_3$ compounds from Johnson – Mattey and Merck. $HfO_2$ was introduced into the melt to obtain $Hf^{4+}$ doped crystals with various contents of 1.0, 3.0, 4.0, 5.0, and 8.0 mol%. In order to obtain single-domain crystals directly during the growth process a dc electric current with a density of about 12 $A/m^2$ was passed through the crystal-melt system. Finally, Y-cut samples were shaped as parallelepipeds and optically polished on all the surfaces.

**Electro-optic measurements**

For EO measurements, we used an experimental setup based on the Sénarmont arrangement associated with a method called *"Time Response Method"* (TRM) [14], which consists into the measurement of the time response of the EO crystal to a fast step voltage. The optical response at short time leads to the clamped coefficient $r^S$, while the optical response at longer time provides the coefficient $r^T$. For an optical arrangement set at the operating point $M_1$ of the transfer function (Figure 1), the so-called linear working point, corresponding to a 50 % transmission i.e. $(I_{max}-I_{min})/2$, is achieved and the time dependence of the variation of the transmitted beam intensity $\Delta i(t)$ is related to the applied voltage $\Delta V(t)$ by:

$$\Delta i(t) = \frac{\Delta n_0^3 L I_0}{\lambda d} r \otimes \Delta V(t) \quad (2)$$

where $\otimes$ is the convolution operator, $I_0 = I_{max} – I_{min}$ represents the total intensity shift of the transfer function, L is the length of the crystal along the beam-propagation direction, d is the sample thickness along the applied electric field direction, $n_0$ is the ordinary refractive index, $\lambda$ is the laser wavelength and r is the EO coefficient involved in the optogeometric configuration defined by the light-propagation and the applied electric field directions.

The frequency dispersion of the EO coefficient can be derived from the ratio of the quantities $\Delta i(\nu)$ and $\Delta V(\nu)$, which are obtained by the *Z* transform of the measured signals $\Delta i(t)$ and $\Delta V(t)$:

$$r(\nu) = \frac{\lambda d}{\pi n_0^3 I_0 L} \frac{\Delta i(\nu)}{\Delta V(\nu)} \quad (3)$$

We have shown [14] that this technique allows to obtain the frequency dispersion of the EO coefficient from DC up to at least 150 MHz, mainly limited by the rising time of the voltage pulse. It is to be mentioned that the values of the coefficients obtained by this technique are absolute values.

## RESULTS

### Electro-optic results

Electro-optic measurements were carried out by means of the TRM method presented above and using a He-Ne laser beam at λ=633 nm. Within experiments devoted to the determination of the EO coefficient $r_{22}$ the beam was propagating along the c-axis of the sample and an external electric field was applied along the a-axis (or equivalently the b-axis). Illustration of experimental data recorded in the 4 mol %-Hf doped LN sample is shown in figure 2.

In this figure we plot the optical signal, Δi(t) (for two different time scales) recorded at the output of the optical setup when the voltage ΔV(t) is applied on the sample. In the long-time range, the optical signal oscillates with a period corresponding to the main piezo-electric frequency resonance. Oscillations do not exist for time shorter than 0.15 µs since the acoustic waves need more time to propagate through the crystal. These oscillations vanish for time larger than several hundreds of microseconds.

The frequency dispersion of the EO coefficient $r_{22}$ is calculated according to Eq. 3, from the ratio of $\Delta i(\nu)$ and $\Delta V(\nu)$ obtained by the *Z* transform of the two time dependent signals. As shown in figure 3.a, the frequency dependence of the EO coefficient $r_{22}$ of the crystal is flat on both sides of the piezo-electric resonances.

The coefficient $r_{22}$ was thus obtained for each crystal under investigation. The dependence of the clamped and unclamped EO coefficients, $r_{22}^T$ and $r_{22}^S$ on the molar hafnium concentration is shown in Figure 4 (see also table I). We can see that both clamped and unclamped EO coefficients $r_{22}$ are slightly smaller than the values in the undoped crystal and do not change with Hf concentration within the experimental error. The low- and high- frequency values of $r_{22}$ in Hf-doped LN are equal to $r_{22}^S$ = 3.7 pm/V and $r_{22}^T$ = 6 pm/V respectively.

This quasi-constant value of $r_{22}$ coefficient versus Hf concentration in LN crystal has to be emphasized since this coefficient presents a very large and non-monotonous dependence for crystals doped with other ions such as Zn or Mg [15-16].

In addition, we have measured the clamped and unclamped EO coefficient $r_c$ and the corresponding dielectric permittivity $\varepsilon_{33}$. Table 1 summarizes the values of the whole set of measurements. The values of the EO coefficient $r_c^T$ obtained versus Hf concentration are in good agreement with those previously measured by Rossella and *al* [13].

**Dielectric measurements**

In ferroelectric inorganic materials, it was established that the EO properties are linked to the linear dielectric properties. In particular, it was shown that the frequency dependence [17] of an EO coefficient reproduces the behaviour of the corresponding dielectric permittivity ε. Such a link between ε and r do still exist in their dependence on the dopant composition. The frequency dispersion of dielectric permittivity $\varepsilon_{22}$ of the 4 mol%-Hf doped LN crystal measured along the c-axis is shown in figure 3.b. As expected, the behaviour of $\varepsilon_{22}$ with frequency is very similar to this of $r_{22}$. We can see that $\varepsilon_{22}$ displays a jump of its value between both sides of the piezoelectric resonances. This difference Δε corresponds to the electromechanical contribution to the static permittivity $\varepsilon^T$, as it will discussed below. We have undertaken the measurements of the dielectric permittivity $\varepsilon_{22}$ as function of frequency for every crystal. The Hf-concentration dependence of the clamped and unclamped permittivities, $\varepsilon_{22}^T$ and $\varepsilon_{22}^S$, are thus reported in figure 5. Values of $\varepsilon_{22}$ in Hf-doped LN crystals differ significantly from those in undoped crystal, but show a weak variation with Hf concentration. All data are reported in table 1.

## DISCUSSION

We discuss the frequency and Hf concentration dependences of the EO coefficients. It is well known that the frequency dependence of the EO coefficients in crystals reflects the various physical processes contributing to the EO effect according to the working frequency. In inorganic crystals the contribution arising from the optical phonons is responsible for a large value at high frequency $r^S$. In piezo-electric crystals, the EO coefficient measured under low-frequency electric field or DC electric field arises from additional contribution related to the crystal deformation via the piezo-electric and the elasto-optic effects. This additional contribution is the so-called acoustic or piezo-optic contribution, denoted $r^a$ and is therefore given by the difference between the unclamped $r^T$ and clamped $r^S$ coefficients.

It can be thus derived from the values of the EO coefficients measured below and above the piezo-electric resonances. Here we found in Hf-doped LN crystals $r_{22}{}^a = r_{22}{}^T - r_{22}{}^S = 2.3 \pm 0.3$ pm/V.

Furthermore, $r^a$ can be estimated from elasto-optic and piezo-electric coefficients, via

$$r^a_{ij,k} = \sum_{lm} p^S_{ij,lm} d_{lm,k} \tag{4}$$

where $p^S$ and d denote respectively components of the elasto-optic at constant strain and of the piezo-electric tensors. According to the point group 3m of LN crystal, we obtain the piezo-optic contribution to $r_{22}$ from Eq. 4 (in Voigt notation)

$$r^a_{22} = -(p_{11} - p_{12})d_{22} + p_{14}d_{15} \tag{5}$$

Using the values of piezo-electric and elasto-optic coefficients, available in the literature [18-19] for the congruent composition only, we found $r^a_{22} = 2.6$ pmV$^{-1}$ which is in a good agreement with the experimental value, within the experimental error (10%.) This piezo-optic contribution is relatively large in LN since it constitutes nearly 40% of the total value $r^T_{22} \sim 6$ pmV$^{-1}$.

Likewise, the difference between the low- and the high- frequency values of dielectric permittivity recorded on both sides of the acoustic resonances can be expressed as:

$$\Delta\varepsilon_{ij} = \varepsilon^T_{ij} - \varepsilon^S_{ij} = \sum_{kl} d_{ij,k} e_{kl} = \sum_{kl} d_{ij,k} C_{kl} d_{l,ij} \quad , \tag{6}$$

where e and C are components of inverse piezoelectric and elastic constant tensors. Therefore the difference $\Delta\varepsilon$ corresponds to the electromechanical contribution to the static permittivity $\varepsilon^T$. For LN (point group 3m) we obtain $\Delta\varepsilon_{22}$ as

$$\Delta\varepsilon_{22} = 2d_{22}^2(C_{11} - C_{12}) - 4d_{22}d_{15}C_{14} + d_{15}^2C_{44} \quad , \qquad (7)$$

We obtained for undoped congruent LN crystal from Eq. 6, $\Delta\varepsilon_{22} = 35$, which is close to the step directly detected in the experiments between both sides of piezo-resonances $\Delta\varepsilon \approx 35$.

From the above consideration, we can interpret the large step between low- and high-frequency values in both the EO coefficient $r_{22}$ and the dielectric permittivity $\varepsilon_{22}$ in LN mainly due to the large piezo-electric coefficient $d_{22}$ and $d_{15}$.

The steps $\Delta\varepsilon$ et $r^a$ are slightly changed when Hf is introduced in the LN lattice. This change can be attributed to a change in piezo-electric and elasto-optic coefficients in doped crystals and generally to the strain effects along a (or b) axis.

By contrast the strain effects on the EO coefficient $r_c$ and the dielectric permittivity $\varepsilon_{33}$ are very small. Indeed we found the $r_c^a$ contribution, $r_c^a = r_c^T - r_c^S = 0 \pm 2$ pm/V which is much smaller than $r_{22}^a$. This contribution can be compared with the calculated value $[p_{33} - (n_0/n_e)^3\ p_{13}]\ d_{33} + [2\ p_{31} - (n_0/n_e)^3\ (p_{11} + p_{12})]\ d_{31} = 0.4$ pm/V [15-16]. This negligeable piezo-optic contribution to the coefficient $r_c$ can be related to the electro-mechanical part of $\varepsilon_{33}^T$ which is smaller than $\varepsilon_{22}$.

## CONCLUSION

We have determined the unclamped and clamped values of the electro-optic coefficients $r_{22}$ and $r_c$ , and the corresponding dielectric permittivity $\varepsilon_{22}$ and $\varepsilon_{33}$ as well, in Hf-doped $LiNbO_3$ crystals with varying Hf concentration. The frequency dependence of the EO coefficient reflects this of the associated permittivity. The piezo-optic contribution of the EO coefficient is much larger in $r_{22}$ than in $r_c$, which is mainly related to the stronger electric field induced –deformation along the a(b) axis, as reflected by the difference between the low-and high-frequency values of $\varepsilon_{22}$

Both EO and dielectric coefficients reveal a weak dependence on the Hf content introduced in the LN lattice, which is attributed to the strain contribution related to the introduction of Hf ions. As the electro-optic properties of LN are preserved by Hf doping, the Hf-doped LN crystals of LN present the advantage if compared to undoped congruent crystal to have a smaller photorefractive damage, especially for concentration larger than the threshold, and therefore should be more suitable for EO and NLO applications.

**ACKNOWLEDGEMENTS**

This work was partially supported by the projects NFSAT-UCEP-02-07 and the grant 95 MSE RA.

**Table 1.** Absolute values of $r_{22}$ and $r_c$ EO coefficients at constant stress ($r^T$) and at constant strain ($r^S$) in Hf:LN obtained at 633 nm and at room temperature and dielectric permittivity $\varepsilon_{22}$ and $\varepsilon_{33}$.

| | $r_{22}^T$ (pm/V) | $r_{22}^S$ (pm/V) | $\lvert r_{22}^T - r_{22}^S \rvert$ (pm/V) | $\varepsilon_{22}^T$ | $\varepsilon_{22}^S$ | $r_c^T$ (pm/V) | $r_c^S$ (pm/V) | $\lvert r_c^T - r_c^S \rvert$ (pm/V) | $\varepsilon_{33}^T$ | $\varepsilon_{33}^S$ |
|---|---|---|---|---|---|---|---|---|---|---|
| Cong. LN | 6.4 ± 0.4 | 3.8 ± 0.3 | 2.6 ± 0.4 | 95 ± 6 | 60 ± 3 | 19.7 ± 2 | 19.7 ± 2 | 0 ± 2 | 35 ± 2 | 30 ± 2 |
| 1%Hf: LN | 5.9 ± 0.4 | 3.6 ± 0.3 | 2.3 ± 0.4 | 87 ± 5 | 76 ± 4 | 19.4 ± 2 | 19.4 ± 2 | 0 ± 2 | 34 ± 2 | 31 ± 2 |
| 3%Hf: LN | 6 ± 0.4 | 3.8 ± 0.3 | 2.2 ± 0.4 | 90 ± 5 | 79 ± 4 | 18.9 ± 2 | 18.9 ± 2 | 0 ± 2 | 35 ± 2 | 31 ± 2 |
| 4%Hf: LN | 6 ± 0.4 | 3.7 ± 0.3 | 2.3 ± 0.4 | 87 ± 5 | 75 ± 4 | 20.3 ± 2 | 20.3 ± 2 | 0 ± 2 | 34 ± 2 | 32 ± 2 |
| 5%Hf: LN | 6.2 ± 0.4 | 3.7 ± 0.3 | 2.5 ± 0.4 | 88 ± 5 | 76 ± 4 | 20 ± 2 | 20 ± 2 | 0 ± 2 | 34 ± 2 | 30 ± 2 |
| 8%Hf: LN | 6.1 ± 0.4 | 3.8 ± 0.3 | 2.3 ± 0.4 | 89 ± 5 | 78 ± 4 | 20.4 ± 2 | 20.4 ± 2 | 0 ± 2 | 36 ± 2 | 33 ± 2 |

**Fig. 1** Optical transmission function of Sénarmont setup versus the angle of the analyzer β. The point $M_0$ is the minimum transmission point for which the output optical signal has a frequency twice the frequency of the applied electric field. $M_1$ is the 50 % transmission point yielding the linear replica of the ac voltage.

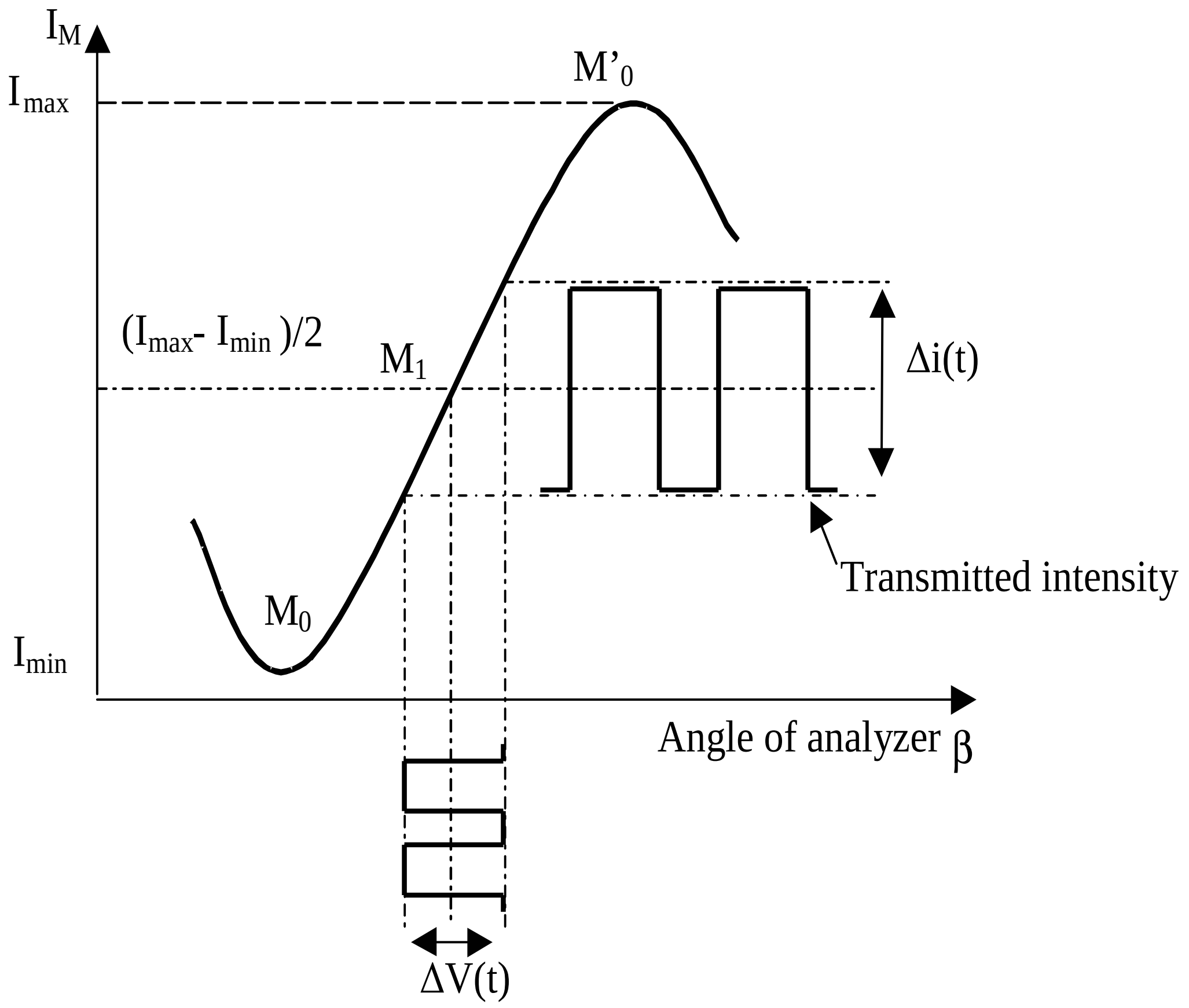

**Fig. 2.** Responses to a step voltage at different time scales in the case of the EO coefficient $r_{22}$ in 4%-Hf doped LN single crystal. Measurements were performed at the wavelength of 633nm.

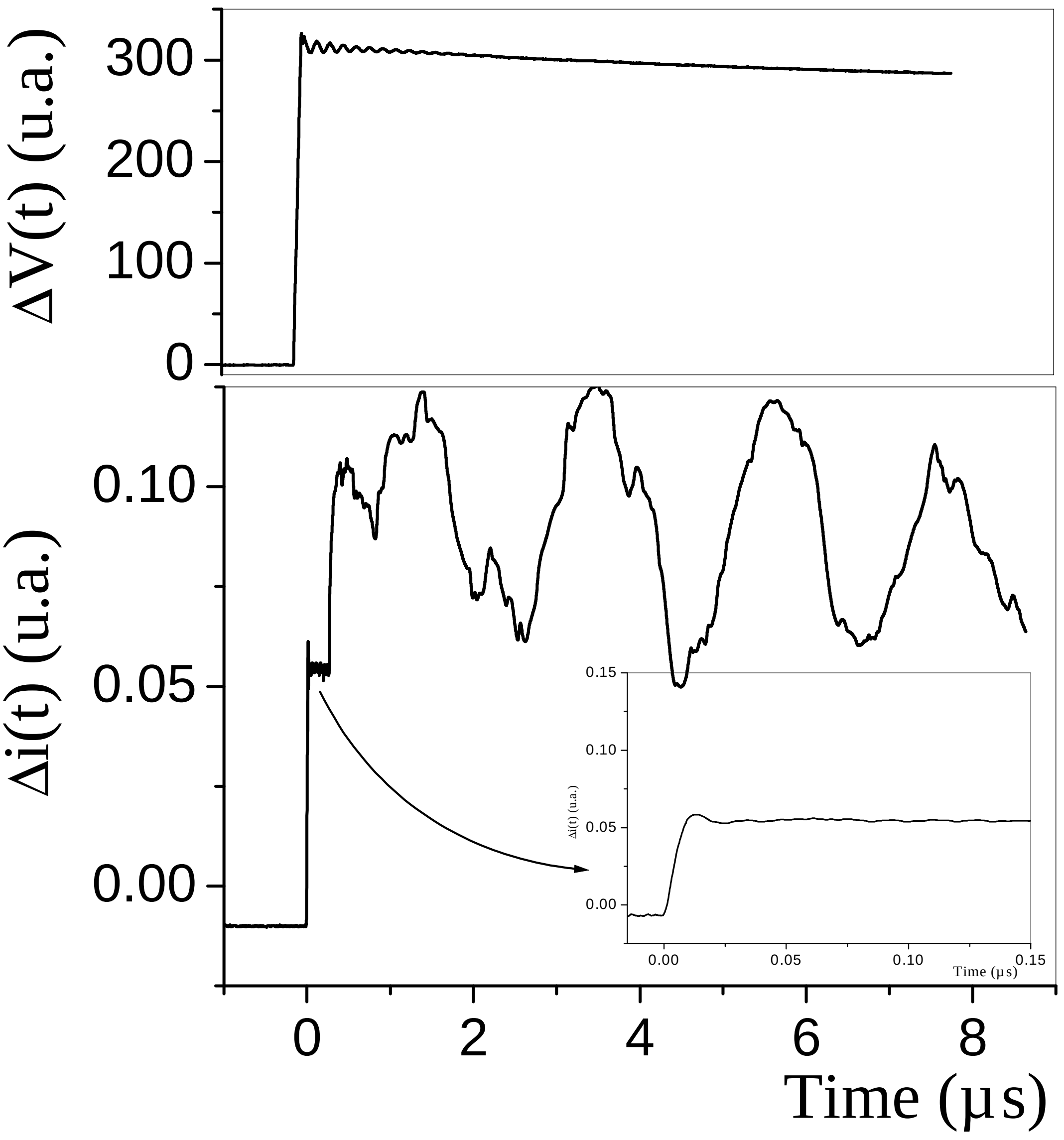

**Fig.3.** Comparison of the frequency dispersions of the dielectric permittivity and of the EO coefficient $r_{22}$ measured in 4 % Hf-doped LN crystal.

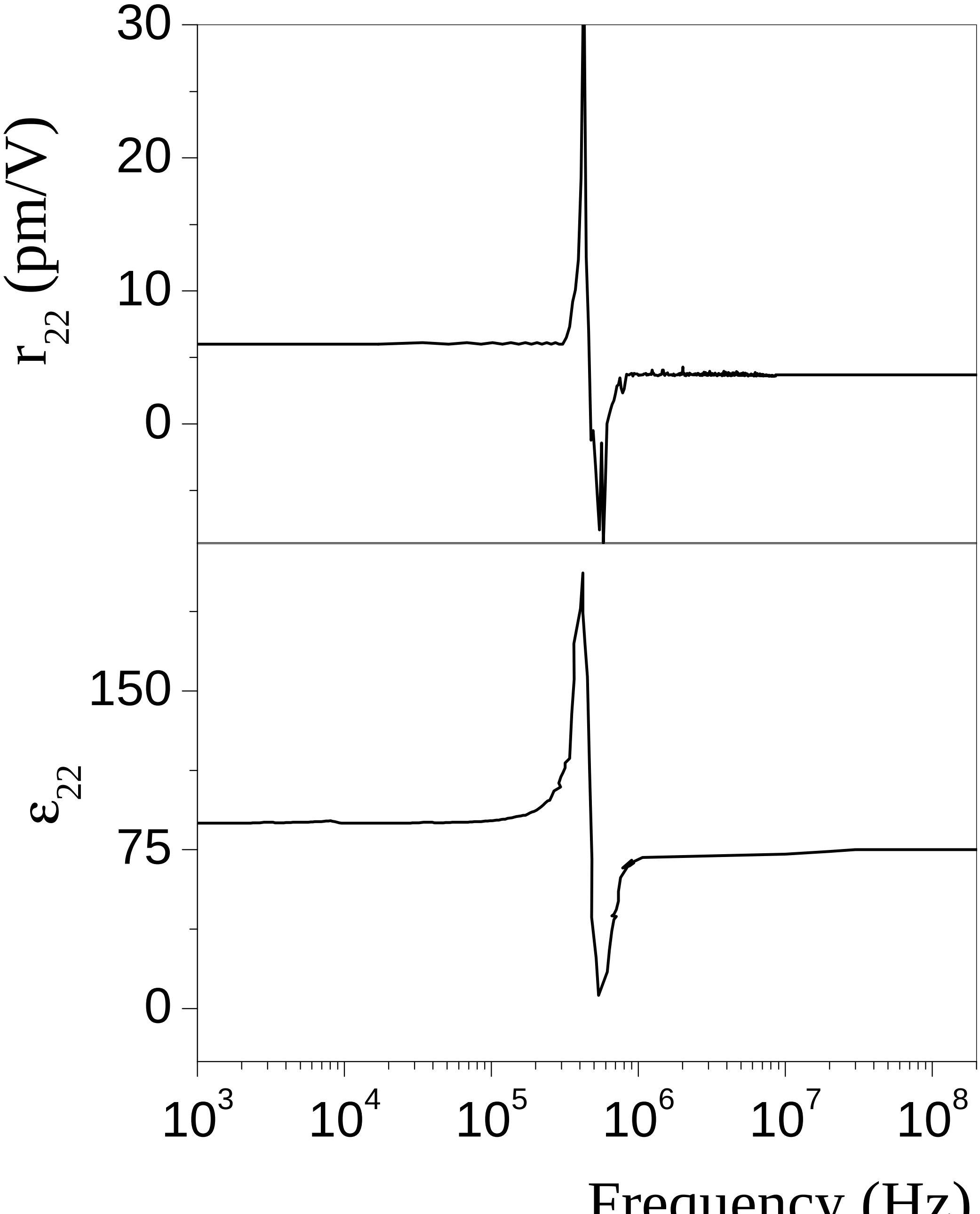

**Fig.4.** EO coefficient $r_{22}$ versus Hf concentration in congruent LN.

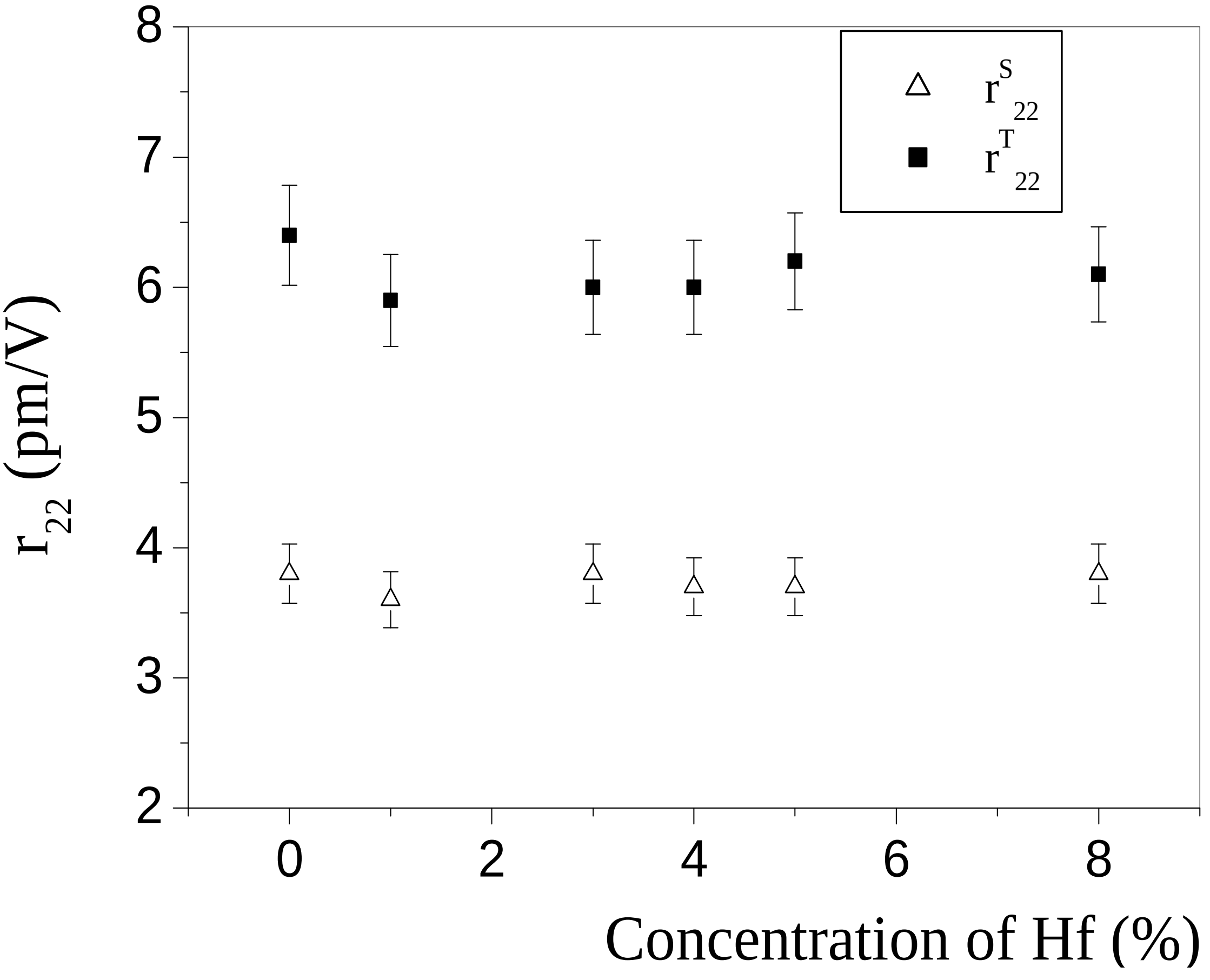

**Fig.5.** Dielectric constant $\varepsilon_{22}$ versus Hf concentration in congruent LN

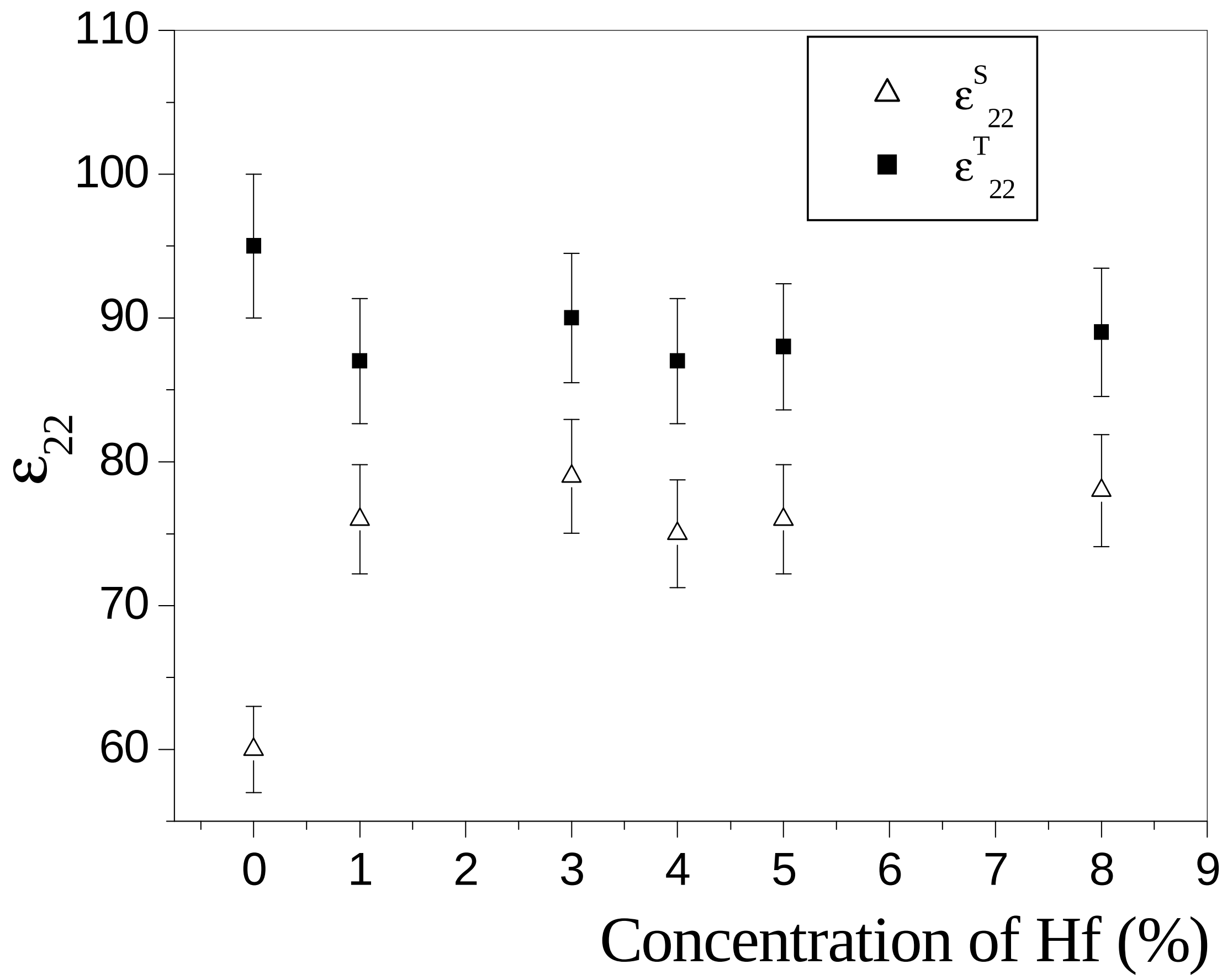